# Application-driven Test and Evaluation Framework for Indoor Localization Systems in Warehouses

Jakob Schyga[1], Johannes Hinckeldeyn[1], Benjamin Bruss[2], Christoph Bamberger[3] and Jochen Kreutzfeldt[1]

[1] *Institute for Technical Logistics, Hamburg University of Technology, Theodor-Yorck-Str. 8, Hamburg, Germany*
[2] *Jungheinrich AG, Friedrich-Ebert-Damm 129, Hamburg, Germany*
[3] *SICK AG, Erwin-Sick-Str. 1, Waldkirch, Germany*

**Abstract**

Despite their potential of increasing operational efficiency, transparency, and safety, the use of Localization and Tracking Systems (LTSs) in warehouse environments remains seldom. One reason is the lack of market transparency and stakeholder's trust in the systems' performance as a consequence of poor use of Test and Evaluation (T&E) methods and transferability of the obtained T&E results. The *T&E 4Log (Test and Evaluation for Logistics) Framework* was developed to examine how the transferability of T&E results to practical scenarios in warehouse environments can be increased. Conventional T&E approaches are integrated and extended under consideration of the warehouse environment, logistics applications, and domain-specific requirements, into an application-driven T&E framework. The application of the proposed framework in standard and application-dependent test cases leads to a set of performance criteria and corresponding application-specific requirements. This enables a well-founded identification of suitable LTSs for given warehouse applications. The *T&E 4Log Framework* was implemented at the Institute for Technical Logistics (ITL) and validated by T&E of a reflector-based Light Detection and Ranging (LiDAR) LTS, a contour-based LiDAR LTS, and an Ultra-Wideband (UWB) LTS for the exemplary applications *Automated Pallet Booking*, *Goods Tracking,* and *Autonomous Forklift Navigation.*

## 1. Introduction

In warehouse environments, everything revolves around the storing and movement of goods. To increase understanding and control of the material flow, position recording is an essential instrument. LTSs enable the determination of the position of various entities, such as people, pallets, and forklifts over time. An examination of the literature reveals the potential of applying LTSs in warehouses. Lee et al. [1] present an application for the tracking of assets in warehouses with Bluetooth technology, while Elser et al. [2] apply the technology for order tracing in single and small batch production. A camera system is applied by Échorcard et al. [3] to determine the absolute position of humans in warehouses. In particular, for navigation tasks in mobile robotics, LTSs represent a main enabler. The localization of robotic forklifts is implemented by Kelen et al. [4] applying odometry data and map-matching algorithms, by Kitajima et al. [5] based on camera systems, and by Beinschob and Reinke [6] on LiDAR-technology. Macoir et al. [7] apply drones for inventory management using UWB.

Despite their potential to increase operational efficiency, transparency, and safety in warehouses the usage of LTSs in logistics environments in practice remains seldom. Potorti et. al. [8] point out the lack of standardized concepts, procedures, and metrics as a reason for low market transparency and consequently low trust of stakeholders, and a slow adoption rate. Morton [9] shares this view and names the definition of application-domain-specific requirements as one of the key open research challenges in the field of indoor localization, alongside the definition of common evaluation methods and metrics. Different frameworks and methodologies for T&E exist but do not satisfyingly fulfill the practical requirements for high transferability, comparability, repeatability, feasibility, and comprehensibility, particularly in the domain of warehouse applications. Depending on the stakeholder's point of view, the aim and requirements of T&E can vary. LTS developers are particularly interested in the characteristics of the system itself and the differentiation and quantification of technical influence factors, such as occlusion or reflection. For applicants and system integrators it is paramount, how the system performs for a certain task in a given environment [8].

This paper examines, in which way T&E of LTSs can be performed systematically to increase the transferability of the results to real warehouse applications. The *T&E 4Log Framework* was designed to address this issue for T&E of LTSs in warehouse environments while satisfying the need for comparability, repeatability, feasibility, and comprehensibility. This is achieved by integrating conventional T&E approaches and the modeling of warehouse environments and processes as well as application-specific requirements into an application-driven T&E framework. The *T&E 4Log Implementation* at the ITL was subsequently developed to validate the proposed framework, based on three LTSs and three warehouse applications.

In the following section, the challenges and the state of the art of T&E for LTSs are discussed. In Section 3 the requirements for the development of the framework are derived based on the stakeholder's perspective, and the *T&E 4Log Framework* is presented. The implementation of the framework is described in Section 4. Section 5 states the specification, execution, and evaluation of experiments. Finally, the results of this work are concluded and an outlook on open research topics is given.

## 2. Discussion of Test and Evaluation of Localization and Tracking Systems

In this section, the challenges of T&E of LTSs are pointed out by discussing the various influencing factors on the localization quality of LTSs. Building on this, literature on the subject of T&E from LTSs is discussed and the need for common methodologies is demonstrated. Existing methodologies are then presented and the special features of the ISO/IEC 18305 International Standard [10] are discussed.

The quality of a system according to the DIN EN ISO 9000:2015 [11] is defined as the degree to which a set of inherent characteristics fulfills requirements. Separating the technical interferences as part of the requirements and LTSs characteristics in this equation, the term localization quality of LTSs describes the system's suitability for a certain application under given interferences regarding the localization. Figure 1 is visualizing these relations.

The characteristics of an LTSs depend on a variety of factors (Figure 1, left), such as the technology (e.g. UWB, LiDAR, accelerometer), the measuring principle (e.g. time-of-arrival, received signal strength), the positioning method (e.g. fingerprinting, pedestrian dead reckoning, multilateration), the implementation (hard- and software), and deployment. For instance, the position accuracy of a UWB-LTS is naturally depending on the position and amount of anchor nodes [12, 13]. Furthermore, the accuracy of a UWB system is influenced by technical interferences, such as occlusion, reflection, or radio interference [14], which in turn are a consequence of the environment. The separation of technical interferences and environment can be transferred to other LTSs and is likewise visualized in Figure 1 (middle). The analysis of the influence of interferences or the environment is subject to several investigations [5, 15, 16], but there is no established method to systematically analyze and compare them or to draw practical conclusions.

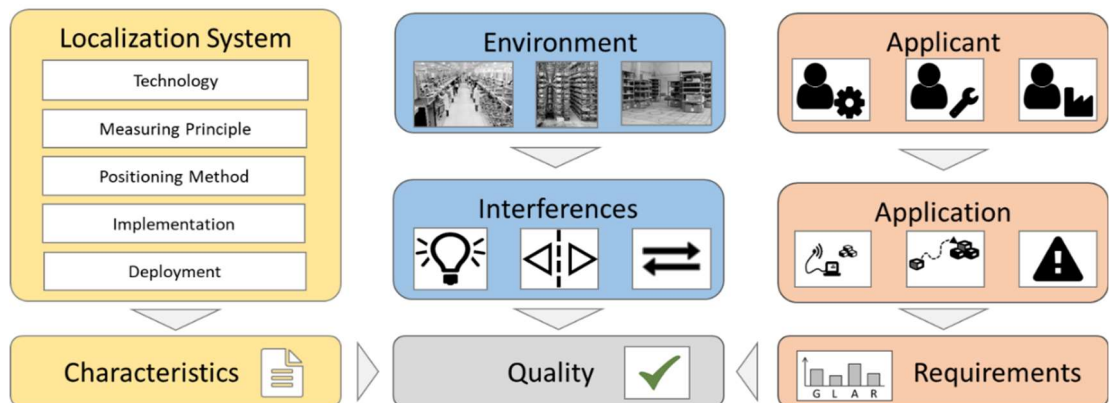

**Figure 1:** Matching requirements and LTS - A multi-dimensional problem.

To determine a system's quality, the characteristics have to correspond to the applicant's requirements. For this reason, many authors have defined partly varying metrics [17–19]. Mautz [19] is giving the most holistic overview of this issue, defining and describing 16 potential user requirements, such as accuracy, latency, and integrity. In addition, Mautz presents a flow chart to capture user requirements and states concrete requirements for selected applications from underground construction to law enforcement. The challenges for performing T&E of LTSs are summarized in the following:

- LTS technologies are highly heterogeneous, ranging from camera systems to inertia sensors.
- The performance of the systems is depending on the deployment, in particular for infrastructure-based systems.
- The physical environment strongly influences the performance of LTSs. The impact of technical influences depends on the system and the environment in turn depends on the application.
- The type and level of requirements vary depending on the applicant and application.

From an applicant's or system integrator's point of view, T&E serves to finding a solution for this multi-dimensional optimization problem [19]. Considerable spatial, financial, and organizational resources as well as extensive knowledge in the areas of measurement technology, statistics, data processing, and evaluation are needed to perform repeatable and transferable testing. Morton [9] points out, that as a consequence, mostly simple proof of concept tests are carried out in practice, which lack comparability, repeatability, and comprehensibility. T&E results are summarized and compared in various surveys [17–20]. Usually, the systems are assigned to an individual technology, measuring principle, or positioning method to draw generalized conclusions.

In particular, as a result of the complex interrelationships between influencing variables, the need for the application of standardized procedures, to increase the results comparability, repeatability, and comprehensibility becomes evident. Different frameworks and methodologies for T&E of LTSs exist, such as the *EvAAL Framework* [21] and the *EVARILOS Benchmarking Handbook* [22]. Both are applied especially in the context of competitions such as the *IPIN Competitions* [23] and the *Microsoft Indoor Localization Competitions* [24]. The *EvAAL Framework* is a set of rules and metrics, particularly designed for use in such competitions. The *EVARILOS Benchmarking Handbook* was a result of the EU FP7 project *EVARILOS – Evaluation of RF-based Indoor Localization Solutions for the Future Internet* [22], focusing on T&E of radio-frequency based LTSs. Building on the findings of *EVARILOS* the ISO/IEC 18305:2016 International Standard – *Test and evaluation of localization and tracking systems* [10] was proposed in 2016 to define terms, test scenarios, performance metrics, and reporting requirements for T&E of generic LTSs. The ISO/IEC 18305:2016 is focused on system testing with a black-box approach, meaning that the LTSs are tested as a whole without consideration of their components or their functionality. In addition, a building-wide testing approach is used. The ISO/IEC 18305:2016 categorizes buildings into five types, such as mines and warehouses. The Entities to be Localized and Tracked (ELTs) are categorized in *Object*, *Person*, and *Robot*, and different mobility modes, such as *backward walking, sidestepping*, etc. are defined. The building types, ELTs, and mobility modes are combined to 14 test scenarios, without consideration of the robot class. Benchmarks of LTSs in real warehouses exist as part of the *PDR Challenge in Warehouse Picking 2017* or the *xDR Challenge for Warehouse Operations 2018* as presented by Ichikari [25]. Since the position estimation in the named competitions is based on a pre-recorded data set, the evaluation is limited to the given systems and configuration. While competitions often lead to a good comparability of the LTS for the defined procedure, the transferability to other applications and the comparability to other benchmarks with differing processes and environments remains limited.

Even though, different methodologies exist, their consistent application in research and above all in the industry is not yet given. The ISO/IEC 18305:2016 currently is the most suitable option, for the matter of determining the localization quality of generic LTSs. However, the ISO/IEC 18305:2016 is the first proposal of a standard for solving the described multi-dimensional optimization problem, and various aspects are still open to debate.

# 3. Application-driven Test and Evaluation of Localization and Tracking Systems

In the previous section, it was shown that the heterogeneous nature of technology and application of LTSs pose various challenges for T&E, which are not satisfactorily addressed in practice. Despite the diverse influencing factors on the localization quality, the practical use of standardized methodologies has been rare. A technology-driven option to reduce the complexity of the described problem is to focus on a specific technology, as with EVARILOS focusing on radio-frequency sensors. From the applicant's point of view, the inner workings of an LTS are not relevant as long as it fulfills all requirements. For instance, if an applicant is aiming to find the optimal LTS to enable an automated pallet booking process, a T&E procedure designed solely for the comparison of radio-frequency-based sensors is misleading. An option more centered on the applicant, is to focus on an application domain. Focusing on a certain application domain reduces not only the complexity of T&E and therefore increases comparability, repeatability, feasibility, and comprehensibility but furthermore enables an application-oriented analysis and therefore increases transferability. Accordingly, the *T&E 4Log Framework* was developed as an application-driven approach for T&E of LTSs in warehouses. In the following, an adapted V-Model is presented, to point out the perspectives of the various stakeholders in T&E. Subsequently, guidelines for the design of the *T&E 4Log Framework* are derived and its current state presented.

## 3.1. Stakeholders and Requirements for Application-driven Test and Evaluation

The requirements on T&E depend on the stakeholder's perspective. In this section, the perspectives of the stakeholders are discussed and requirements, depending on the stakeholder and the phase of the T&E process are derived. Furthermore, concrete design guidelines for the development of an application-driven T&E framework for LTSs in warehouses, with the aim of finding a suitable compromise in the field of tension between transferability, comparability, repeatability, feasibility, and comprehensibility, are presented.

The V-Model, commonly applied as a management tool for software development and software tests was adapted to fit an application-driven T&E process for LTSs as depicted in Figure 2. Depending on the phase in the T&E process, the presented V-Model visualizes the perspectives, functions, and requirements of applicants, LTS developers, and test staff. Applicants define the application including the processes, environment, and requirements. They aim for selecting the optimal LTS for their specific needs. It is required that the results of the various LTSs are comparable to other LTSs, environments, and processes (**Comparability**) and that the results can be reproduced in real scenarios (**Transferability**). The system developers have to configure and deploy the LTS according to the needs of the applicants and therefore ensure, that the results fulfill the requirements for comparability and transferability. The T&E results can be used by the system developer to improve the LTS, its configuration, or its deployment. **Repeatability** is the ability to reproduce the same results for testing the same LTS in the same environment [19] and is a precondition for comparability. The test staff, in turn, defines a concrete test setup and scenarios, performs the experiments, processes the data, and proves the validity of the results, while ensuring repeatability, comparability, and transferability. **Feasibility** measures the required spatial, financial, organizational, and personnel effort to perform the T&E of LTSs. Naturally, performing the defined test cases has to be feasible. In addition, the **comprehensibility** of the processes and the data has to be given at every stage of the V-Model for the respective stakeholders.

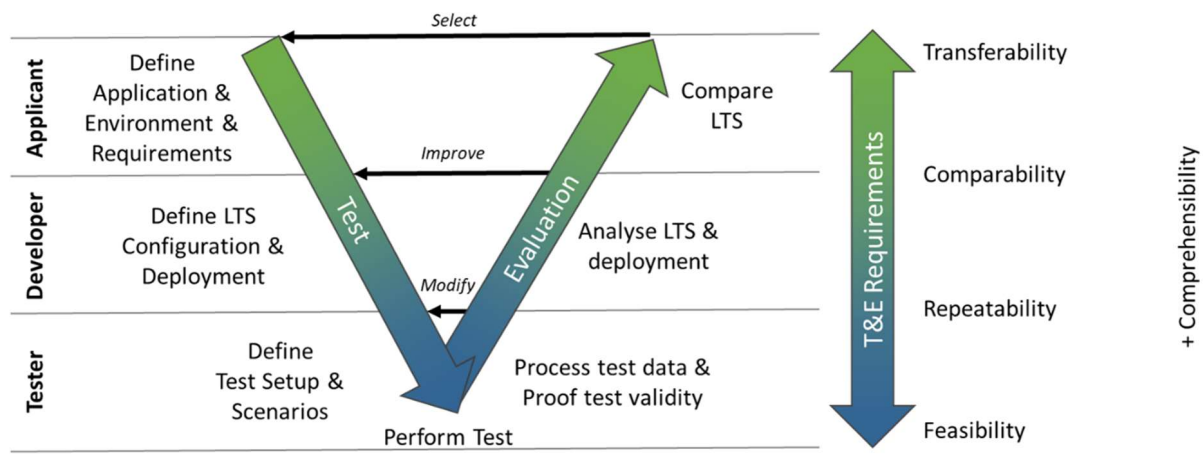

**Figure 2:** The V-Model - Stakeholders, functions and requirements.

The following guidelines (*and main objectives*) for the design of the *T&E 4Log Framework* were derived from the analysis of existing specifications and benchmarks in the literature as well as requirements for the T&E of LTSs from the industry, under consideration of the presented V-Model.

1. Definition of domain-specific evaluation metrics. (*Transferability*)
2. Possibility of application-dependent T&E. (*Transferability*)
3. Possibility of the evaluation of LTSs for mobile robotics. (*Transferability*)
4. Systematic evaluation for the LTS's performance depending on the environment. (*Transferability*)
5. Openness for LTS Technology. (*Transferability & Comparability*)
6. Integration of standardized test scenarios. (*Comparability & Repeatability*)
7. Integration of terms, metrics, processes, and considerations from the ISO/IEC 18305 Standard. (*Comparability, Repeatability & Comprehensibility*)
8. Focus on warehouse applications. (*Feasibility*)
9. Separation of a focused application evaluation designed for applicants and holistic performance evaluation for system developers and test staff. (*Comprehensibility*)

The main focus in the development of the framework is on the transferability of the T&E results for warehouse scenarios to increase the informative value of the results for real scenarios. However, regarding Figure 2 it is evident that application-driven T&E cannot be performed successfully without consideration of the system developer and test staff, their functions, and requirements.

### 3.2. The T&E 4Log Framework

In the following, the design guidelines presented in the previous section are transferred into a T&E framework, under consideration of the requirements for transferability, comparability, repeatability, feasibility, and comprehensibility, by integrating established methodologies, guidelines, and tools. First, an overview of the main characteristics of the T&E 4Log Framework is stated. Next, the framework architecture is presented and the functionalities of the individual modules are explained.

As the LTSs are evaluated independent of the technology, A system-level, black-box approach is chosen, to ensure the evaluation of the LTS. The *T&E 4Log Framework* can furthermore be categorized as repeatable and building-wide testing according to the ISO/IEC 18305. To ensure repeatability, a reference deployment is defined for each technology. Environmental influences will be recorded, according to the reporting requirements of the ISO/IEC 18305. The *T&E 4Log Framework* enables the data evaluation at chosen test points as well as the utilization of all LTS pose measurement data. Depending on the considered performance metric, the appropriate data source is chosen. For instance, position accuracy can be evaluated with higher repeatability based on the evaluation at predefined evaluation poses, while the latency can be calculated with higher accuracy by applying the full data source. To enable the full utilization of the position measurement data, a reference LTS is required as a Ground Truth (GT). As recommended in the ISO/IEC 18305, the accuracy of the GT has to be at least one order of magnitude better than the accuracy of the LTS under test, and between 50 and 100 evaluation poses should be chosen. Tolerances are to be defined to pass predefined evaluation poses with a satisfying path accuracy. To enable the analysis of LTS for mobile robotics, such as forklifts or drones, the framework, and the GT have to enable the T&E of LTS with up to six degrees of freedom (6-DoF).

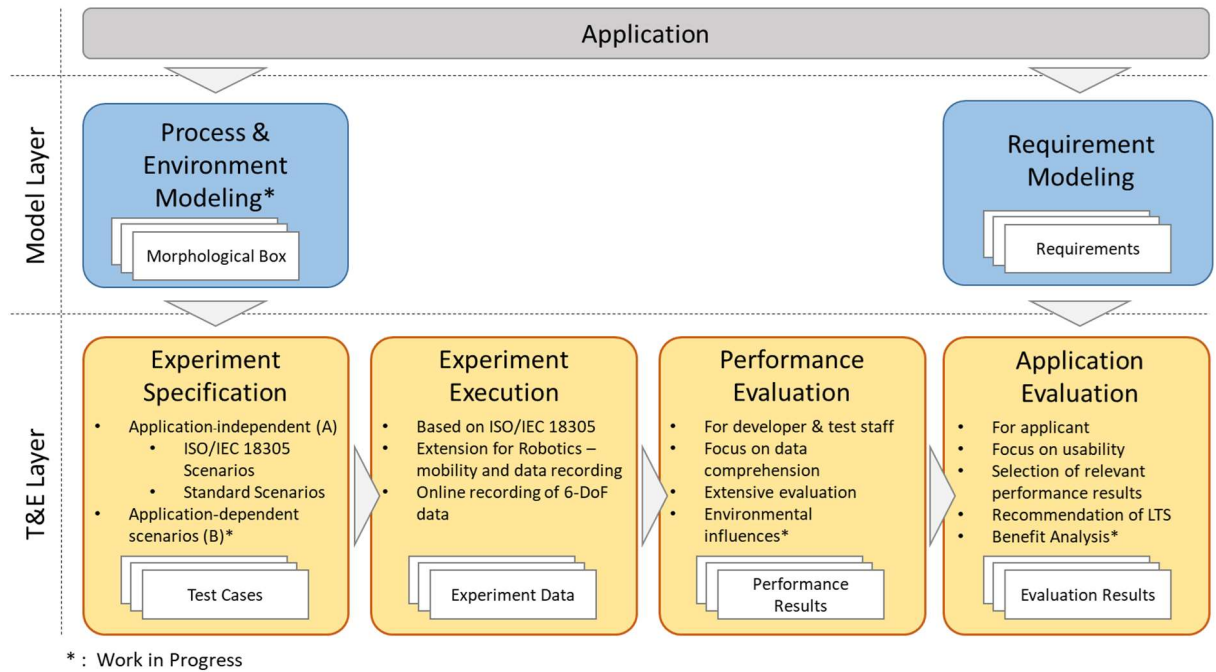

**Figure 3:** Overview of the *T&E 4Log Framework*.

The T&E 4Log Framework contains six function modules with defined interfaces as visualized in Figure 3. Model *Layer* consists of the modules *Process & Environment Modeling* and *Requirement Modeling*. The *Model Layer* serves to abstract a practical application and to generate defined handover documents that influence the subsequent T&E procedure. The modules *Experiment Specification*, *Experiment Execution*, *Performance Evaluation*, and *Application Evaluation* build the *T&E Layer* of the framework. The specifications from the ISO/IEC 18305 are integrated into the modules *Experiment Specification*, *Experiment Execution,* and *Performance Evaluation,* and extended by additional specifications. The *Morphological Box, Requirements, Test Cases, Experiment Data, Performance Results,* and *Evaluation Results* are the handover documents of the respective module and form the basis for the following modules.

In the *Experiment Specification* the trajectory, the ELT, the testbed, the GT, and the configuration of the spatial environment are defined in a *Test Case*. Furthermore, the definition of all parameters according to the reporting requirements from the ISO/IEC 18305 are included. *Scenarios* define a set of parameters for a *Test Case* and are independent of a certain testbed in principle. *Scenarios* are divided into the categories application-independent (**A**) and application-dependent (**B**). Due to the complex superposition of various errors, standard scenarios are proposed. In addition to the scenarios suggested by the ISO/IEC 18305, the following *Scenarios* of **Category A** are defined, to enable the determination of certain performance metrics with high repeatability:

1. T&E 4Log Standard Dynamic-Scenario – This scenario defines a realistic trajectory over the entire test area with several curves and moderate velocity of the ELT, comparable to walking pace.
2. T&E 4Log Standard Static-Scenario – This scenario includes a similar trajectory to the previous one, but with the additional condition of standing still at the evaluation poses. The LTS measurement will only be recorded successfully if a predefined velocity threshold is undershot. This enables the differentiation between velocity-dependent errors, such as time offsets and velocity-independent errors.
3. T&E 4Log Repeatability-Scenario – Repeatability is defined as the measurement precision under a set of repeatability conditions, meaning the closeness of measurement at evaluation poses [26]. Useful repeatability conditions can be a horizontal or spatial position or pose inside a given tolerance. This scenario is suggested for the determination of the LTS performance, regarding repeatability under different conditions.

4. *T&E 4Log Latency-Scenario* – Mautz [19] defines system latency as the delay with which the requested information is available to the consumer of information. Not all LTSs provide their position data with a timestamp and even if, the timestamp cannot be trusted without restriction. The calculation of the latency is performed based on the position error. The position data is provided with a timestamp at the time of arrival at the consumer of information. The systematic, velocity-dependent error of the timestamped data, combined with the velocity at this position provided by the GT is used to calculate the system latency. If the scattering of the LTS' position signal is too high in comparison to the ELT's velocity, the latency cannot be determined with satisfactory quality. The T&E 4Log Latency-Scenario addresses this by provoking velocity-dependent errors and minimizing other systematic errors.
5. *T&E 4Log Coordinate Alignment-Scenario* – From the application point of view, it can be interesting to determine how accurate an LTS has been calibrated and aligned with other coordinate systems. If the mean pose errors are simply eliminated as a bias, as suggested by Potorti [8], the quality of the alignment is disregarded. In the *T&E 4Log Framework,* a scenario is proposed, in which the ELT is moved evenly across the entire test area with static measurements at the evaluation poses. The results from this scenario can be used to align the coordinate systems of the GT and the LTS retrospectively and evaluate the alignment quality of various methods, applying the least-squares-method.

Besides ELT and trajectory information, *Scenarios* can contain specifications of the LTS configuration, deployment, or Environment configuration. *Scenarios* of **Category B** can be defined corresponding to real-world applications, based on the *Process & Environment Modeling* module. An extendable morphological box for the abstraction of warehouse applications is defined to transfer a real-world application into a *T&E 4Log Scenario* of **Category B** and subsequently into a *Test Case*. At the current state of work, the finalization and testing of this module are still in process and therefore not presented in this work.

Experiments are performed in the *Experiment Execution* according to the defined *Test Cases* in the *Experiment Specification*. Functional extensions to the ISO/IEC 18305 International Standard are:
- Possibility of manual or automatic mobility with up to 6-DoF
- Online data stream of up to 6-DoF GT data, LTS data, and ELT control.
- Additional conditions for path accuracy and measurements, such as the ELT's velocity, or the LTS measurements at predefined poses.
- Online ELT control.

Potorti [8] criticizes metrics defined in the ISO/IEC 18305, which are difficult to interpret for the applicant and calls for the restriction to the 95 %-quantile to increase comprehensibility of the data. As the applications of LTSs vary widely, it is not satisfactory for the applicant to focus on the 95 %-quantile. For instance, in the case of simple indoor goods tracking, it may be sufficient to consider the 95 %-quantile with moderate requirements for functional reliability. For the safety-critical application of an autonomous forklift truck, the observation of the 99.99 %- quantile would be more meaningful. To resolve this conflict between simplicity, comprehensibility, and transferability of the results, the *Performance Evaluation* in the *T&E 4Log Framework* is separated from the *Application Evaluation*. For *Scenarios* of **Category A,** the *Performance Evaluation* can be carried out independently from a certain application. The *Performance Evaluation* aims to obtain a holistic understanding of the *Experiment Execution*, the *Experiment Data,* and thus the characteristics of the LTS. The module is therefore expressly not aimed at the applicant but the test staff and LTS developer. Performance metrics include the proposed metrics from the ISO/IEC 18305 and some additional ones, such as the *clock offset,* the *mean orientation error,* or the *drift of position error over time.* The determination of interferences from the test environment is a supplementary function, which is not yet fully defined and validated.

The *Application Evaluation* is based on the requirements derived from the *Requirement Modeling* and on the *Performance Results* to determine the localization quality of an LTS. In the *Requirement Modeling,* several guidelines are defined to derive requirements for an application to be specified, which correspond to the performance metrics. For instance, the guidelines contain an estimate of the required update rate, based on different criteria. For the dynamic case, the decisive criterion is the maximum position error at the ELT's maximum velocity, while for static cases the maximum time difference between two updates is considered. Requirements can be classified as 'shall' or 'must' and supplemented by a benefit value for optional benefit analysis. The direct comparison between corresponding performance metrics and requirements enables the recommendation of LTSs for an application. Table 1 defines the default requirements and assigns them to the corresponding *T&E 4Log Framework* performance metrics.

**Table 1**
Definition of the *T&E 4Log Frameworks* default requirements and corresponding performance metrics.

| Requirement/ Metric | Requirement definition | Performance metric |
|---|---|---|
| Functional reliability | The ability of an observation item to fulfill a required function under given conditions for a given time interval. [27] | Sets a necessary condition for all other scattering functional parameters of an LTS by determining the quantile to be considered. |
| Absolute position accuracy | Degree of closeness of an estimated or measured position at a given point in time with the true value. [26] | Euclidian error distance between estimated position and GT position at the predefined evaluation poses. |
| Orientation accuracy | Degree of closeness of an estimated or measured orientation at a given point in time with the true value. [26] | Absolute orientation error of estimated orientation and GT orientation at the predefined evaluation poses. |
| System latency | Delay with which the requested information is available to the user. [19] | Average Euclidian position offset in direction of ELT's movement between the GT and the LTS, divided by the velocity for every LTS measurement with a velocity over a predefined threshold. |
| Update rate | The frequency with which the positions are calculated on the device or an external system. [19] | Total test time divided by the number of position measurements of the LTS. (For periodic updates) |

The separation of the T&E process into the described modules, with the defined interfaces, enables the test staff to design simple test cases based on the ISO/IEC 18305 with high feasibility and repeatability as well as application-dependent test cases with high relevance for an applicant. The extension of features from the ISO/IEC 18305 increases the options for implementation, such as the T&E of LTSs for mobile robots or drones. The issue of transferability of test results is addressed by analyzing and abstracting the application as part of the *Model Layer* and subsequent use of the results in the *T&E Layer*. The transferability is furthermore increased by the concrete assignment of requirements to performance metrics, while the definition of standard scenarios enables good comparability. The separation of the *Performance Evaluation* from the *Application Evaluation* ensures comprehensibility without loss of information for the system developers and test staff.

## 4. Reference Implementation

The reference implementation of the *T&E 4Log Framework* at the *ITL*-Testbed is described in this section, to demonstrate the feasibility of the framework and to examine its validity. Furthermore, the implementation can be used as a reference for researchers to put the *T&E 4Log* or similar frameworks into place. The description of the reference implementation is not taking the modules or functions into account which are marked as 'Work in Progress' in Figure 3. First, the hardware and then the software components are explained. An overview of the *T&E 4Log Implementation* is given in Figure 4.

The **Environment** within the hall of the ITL is set up and specified by the test staff. The test area has a rectangular base area of 100 m$^2$ and can be equipped with various objects such as partition walls, aisles, pallets, etc. The **Mobility** of the LTSs can be implemented through various ELTs, including a person, an automated guided vehicle, a manually operated or automated forklift truck, and a manually operated handcart. As a **Ground Truth**, an optical passive motion capture system similar to the one examined by Hansen et al. [28] was put into operation at the *ITL-Testbed*. The system of the manufacturer *Qualisys* enables the determination of the 6-DoF pose and the velocity of the ELT with an absolute position accuracy of less than 1 mm. According to the recommendation from the ISO/IEC 18305, this enables the examination of LTSs with an absolute position accuracy of up to 10 mm. The pose data is transferred online to the *Experiment Execution* using the Robot Operating System (ROS), while the experiment is carried out. The GT data is also used by the ELT to ensure high path accuracy and repeatability. The **LTSs** to be examined are deployed and measured according to the reference configuration for the respective technologies. The deployed LTSs are described as part of the validation in the following section. If the LTS does not provide a ROS interface, a simple interface must be written. The alignment of the coordinate axes between the LTSs and GT as well as between the localization device and the ELT can either be carried out manually in advance of the Experiment Execution or minimized afterward based on the *T&E 4Log Coordinate Alignment*-Scenario. The time synchronization between LTSs and GT is performed via the Precision Time Protocol defined in the IEC 61588.2009 [29] with a verified offset of less than 0.5 ms.

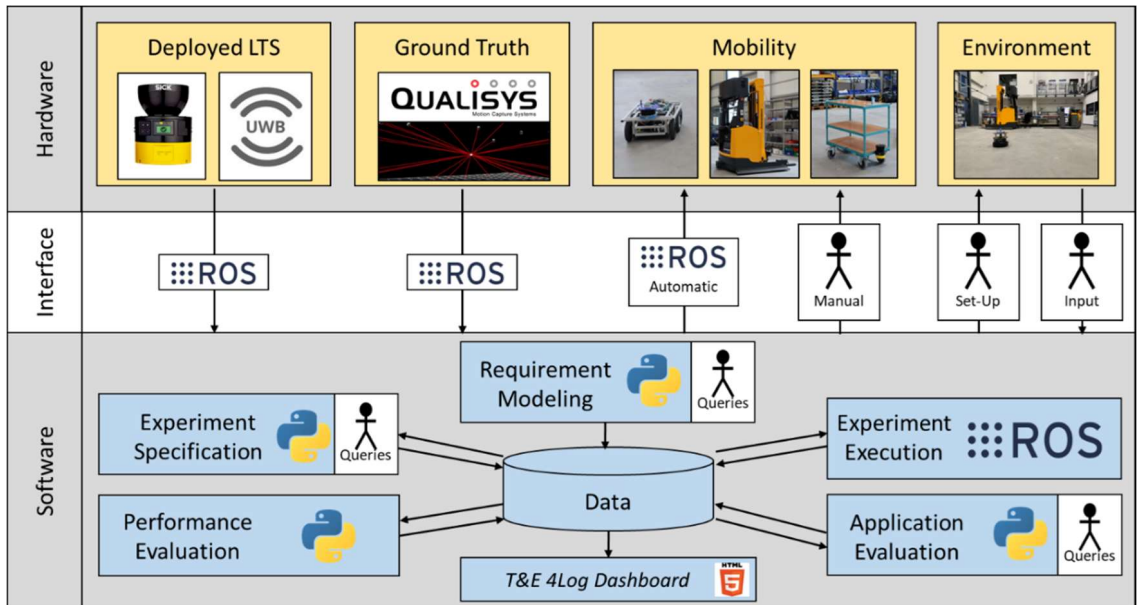

**Figure 4:** Implementation of the *T&E 4Log Framework* at the ITL.

On the software side, a separate code was developed for each T&E module of the *T&E 4Log Framework*. Logical functions are implemented in Python3 and ROS-Nodes in C++ and Python2. The **Experiment Specification** requires a set of user queries to systematically define the *Test Cases* and record the reporting requirements. The result is a list of poses to define the trajectory and a set of information, conditions, and thresholds required for the *Experiment Execution* and the *Performance Evaluation*. Furthermore, requests for the manual setup of the test environment are generated. The **Experiment Execution** handles the data recording and the automatic or manual routing of the ELT along the evaluation poses. The raw LTS and GT pose data from the ELT are saved in a separate ROS-bag file for each experiment. As part of the **Performance Evaluation,** the GT measurement data is interpolated and the LTS measurement data, corresponding timewise to the smallest difference between GT pose and evaluation pose, is extracted. Various performance metrics are subsequently calculated. The **Requirements Modeling** requires a set of user queries, leading the user to the determination of requirements, based on the guidelines from the *Requirement Modeling*. Finally, in the **Application Evaluation,** the requirements are matched to the performance metrics according to Table 1.

The *Test Cases*, *Experiment Data*, *Performance Results*, and *Evaluation Results* are saved in YAML files, to enable good readability. The *T&E 4Log Dashboard* was developed to present the *Experiment Specification*, the *Experiment Data,* the *Performance Results,* and the *Evaluation Results,* including various interactive graphics to increase the comprehensibility for the test staff and LTS developer.

## 5. Validation

For the validation of the *T&E 4Log Framework*, a UWB-based system and a reflector-based, as well as a contour-based LiDAR system were tested and evaluated based on exemplary warehouse applications. The localization of the UWB LTS LOCU from SICK AG is based on the time difference of arrival measurements as described by Morton [9]. Both LiDAR systems consist of the same multilayer LiDAR scanner (microScan3, SICK AG) for the emission and detection of laser impulses, control unit (SIM1000, SICK AG), and the software (LiDAR-LOC, SICK AG) for the self-localization. The pose of the sensor is calculated by comparing scan points with a prerecorded map based on a particle filter [30]. The main difference between the LiDAR LTSs is that the reflector-based system includes the known position of reflectors, in addition to the contour of the spatial environment. The LTSs were deployed at the *ITL-Testbed* according to the technology-specific reference implementation. As the *Process & Environment* module is currently being finalized, the *Scenarios* of **Category A** are applied without the consideration of environmental influences. The *Application Evaluation*, however, is carried out application-dependent based on the applications *Automated Pallet Booking*, *Goods Tracking*, and *Autonomous Forklift Navigation*. In this section, the process of going through the modules of the *T&E 4Log Framework* is described for the *T&E 4Log Standard Dynamic-Scenario*. The results of the process are a list of performance metrics for the three LTSs and a recommendation of LTSs for the applications.

### 5.1. Experiment Specification and Experiment Execution

The manually guided handcart (Figure 5, left) was selected as ELT for the experiment. The handcart is equipped with a UWB-tag and the micoScan3 of the respective LTSs. Additionally, the handcart is equipped with passive GT reflectors for the GT localization of the ELT. The environment at the *ITL-Testbed* is set up in its basic configuration without any static or dynamic obstacles inside the test area. The scheme in Figure 5 (right) is presenting the test setup, including the environment, the defined trajectory, the positions of the LiDAR reflectors, the UWB anchors, and the GT cameras. Four LiDAR reflectors and four UWB anchors are installed in the hall at regular intervals while avoiding symmetries. The coordinate alignment between the LTSs and the GT was carried out by determining the reflectors' respectively the anchors' position in the GT coordinate system. The spatial environment is characterized by logistics objects, such as a pallet shelf, shelving racks and a conveyor belt. The trajectory consists of 63 evaluation poses. The specified experiment was performed for the three LTSs by manually guiding the handcart along the trajectory. Each experiment was carried out three times to check for repeatability.

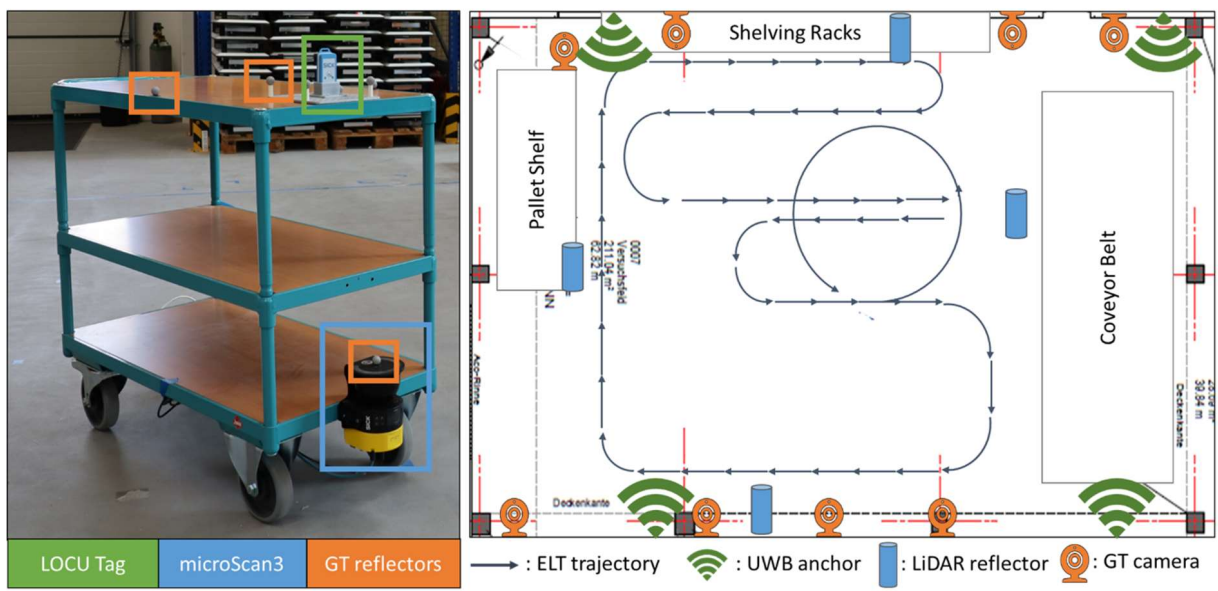

**Figure 5:** Handcart with localization devices (left); Scheme of test setup and trajectory (right).

## 5.2. Performance Evaluation

In the following, the *Performance Evaluation* based on the *Performance Results* of the experiments for the reflector-based LiDAR is exemplarily discussed. The experiment shows good repeatability with an absolute mean horizontal position error of 22.4 mm, 24.7 mm, and 20.0 mm for the three experiments of the same *Test Case*. Figure 7 presents an error scatter plot (left) depending on the ELT's orientation and the horizontal Euclidean position error as a function of measurement time, depending on the ELT's velocity for one experiment.

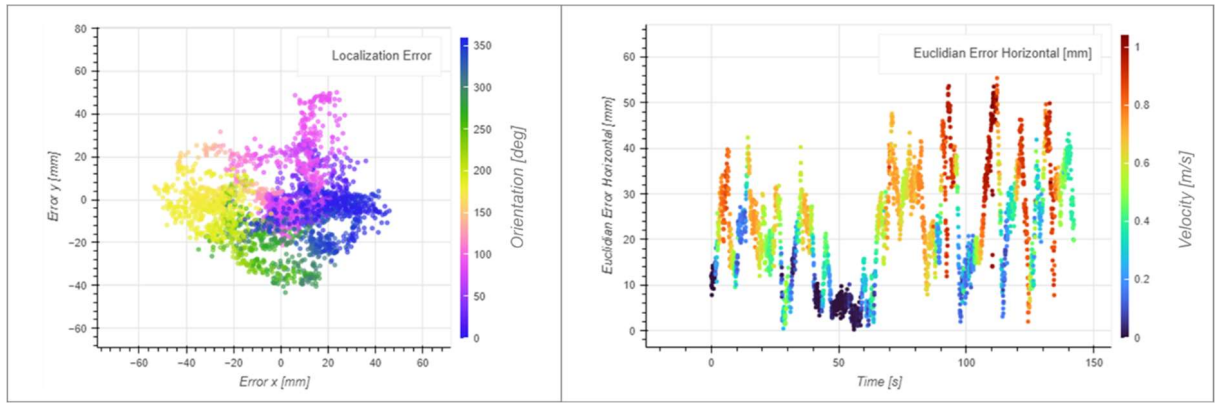

**Figure 7:** Extract from the *T&E 4Log Dashboard*: Error scatter (left), absolute horizontal position error over time (right).

Systematic errors, depending on the ELT's orientation and velocity exist. A systematic error solely based on the orientation could be caused by a poor misalignment of the localization device with the GT. Since the velocity also correlates with the error, a time offset between the LTS and GT measurement is a plausible explanation. Since the GT is synchronized with the LTS with high accuracy by the Precision Time Protocol, the cause is to be expected in the LTS. As the source of the error does not lay in the T&E process the experiment results remain valid. This is to show the importance of carefully checking the data and the relevance of an interactive dashboard to support developers and test staff to prove the validity of the test results.

Table 2 shows a selection of performance metrics (mean and standard deviation) for one experiment of the same *Scenario* for each of the deployed LTSs. The position accuracy of the reflector-based LiDAR system and the contour-based LiDAR system are in a similar range, of a few centimeters. The LiDAR systems additionally generate orientation data. For the orientation as well as the horizontal accuracy the reflector-based system shows a slightly better performance. In the given configuration, none of the deployed LTSs generates vertical position data. The UWB system shows particularly high values for the mean position error in x-direction. The accuracy of the UWB system could therefore be significantly improved by optimizing the coordinate alignment between LTS and GT. It becomes evident, that the various errors can have a variety of causes and correlations. Therefore, it is important to proceed methodically, differentiate between various systematic errors, and prove the validity of the results.

**Table 2**

Selected performance metrics (mean ± standard deviation) from the *Performance Evaluation*.

| Performance metric | Unit | LiDAR - Reflector | LiDAR - Contour | UWB |
|---|---|---|---|---|
| Absolute mean horizontal error | [mm] | 22.3 ± 11.8 | 62.1 ± 26.7 | 185.2 ± 11.8 |
| Absolute mean vertical error | [mm] | - | - | - |
| Absolute mean orientation error | [deg] | 0.9 ± 1.7 | 1.31 ± 2.6 | - |
| Mean position error x | [mm] | -0.16 ± 19.8 | -12.6 ± 32.6 | 91.8 ± 106.8 |
| Mean position error y | [mm] | -2.6 ± 15.5 | 50.5 ± 28.2 | -21.1 ± 160.4 |
| Mean orientation error | [deg] | 0.7 ± 0.7 | 0.75 ± 1.05 | - |
| Update rate | [Hz] | 20.4 | 20.4 | 8.2 |

## 5.3. Requirement Modeling & Application Evaluation

In this section, the *Requirement Modeling* and the *Application Evaluation* are demonstrated. The warehouse applications and the considerations to be made for determining LTS's requirements according to Table 1 are presented in the following. The requirements are exemplarily quantified.

**Goods Tracking** – *Tracking of goods movements to analyze the material flow in a warehouse.* The position of the goods has to be assigned to the shelf aisle. This condition is mapped to an absolute position accuracy of 1 m. Moderate requirements regarding the functional reliability of the localization lead to the consideration of the 95%-quantiles. Concerning the update rate and the system latency, low requirements of 0.1 Hz and 10 s are chosen.

**Automated Pallet Booking** – *Pallets are automatically checked in and out of the warehouse management system when they are stored and retrieved from the shelf using a forklift, based on the position of the fork.* Incorrect bookings in the WMS lead to high follow-up costs and must therefore be avoided. The high functional reliability of the localization of 99.9% is therefore considered. To assign the pallet to be stored or retrieved to the corresponding compartment in the pallet rack, both the horizontal and the vertical position of the fork must be determined with the corresponding accuracy. This is converted into an absolute accuracy requirement of 200 mm horizontally and 500 mm vertically. The orientation of the forklift must also be determined enabling the reliable identify the shelf side, resulting in an absolute orientation requirement of 30°.

**Autonomous Forklift Navigation** – *Forklift, that can independently follow a global path and a local path around obstacles.* Due to the safety-critical function of the forklift's localization in complex scenes and mixed traffic, high demands are made concerning all requirements of absolute horizontal accuracy and orientation accuracy. Considering a maximum velocity of the forklift of 2 m/s and a maximum delay distance of 200 mm yield in a system latency requirement of 100 ms.

The requirements are compared with the corresponding performance metrics as part of the *Application Evaluation*. Table 3 is giving an overview of the requirements and indicates, which LTS fulfills the requirements. All requirements must be met by an LTS for its overall suitability for an application. According to the results of the *Performance Analysis* and the *Requirement Modeling,* the requirements for the *Goods Tracking* are met by all the examined LTSs. The *Automated Pallet Booking* is the only application with a requirement for vertical position accuracy. As none of the tested LTSs provide information about the vertical position, no suitable LTS is found. The requirements from the application *Autonomous Forklift Navigation* are only fully met by the reflector-based LiDAR, as the others do not provide the horizontal position data with an accuracy of less than 50 mm and a functional reliability of minimum 99.99 %.

**Table 3**

Results from the *Requirement Modeling* and *Application Evaluation* of the tested LTSs (U: UWB, R: LiDAR reflector-based, C: LiDAR contour-based) for selected Applications. ($Q_X$: X % - quantile)

| Evaluation Criteria | | Goods Tracking | Automated Pallet Booking | Autonomous Forklift Navigation |
|---|---|---|---|---|
| Absolute position accuracy horizontal | Requirement | $Q_{95}$ < 1000 mm | $Q_{99.9}$ < 200 mm | $Q_{99.99}$ < 50 mm |
| | Suitable LTSs | U, R, C | R, C | R |
| Absolute position accuracy vertical | Requirement | - | $Q_{99.9}$ < 500 mm | - |
| | Suitable LTSs | U, R, C | - | U, R, C |
| Absolute orientation accuracy | Requirement | - | $Q_{99.9}$ < 30° | $Q_{99.99}$ < 4° |
| | Suitable LTSs | U, R, C | R, C | R, C |
| System latency | Requirement | < 10 000 ms | < 1000 ms | < 100 ms |
| | Suitable LTSs | U, R, C | U, R, C | U, R, C |
| Update rate | Requirement | > 0.1 Hz | > 1 Hz | > 20 Hz |
| | Suitable LTSs | U, R, C | U, R, C | R, C |
| Overall Suitable LTSs | | U, R, C | - | R |

## 6. Conclusion and Future Work

High heterogeneity of LTSs and their application, in combination with strong influences on the localization quality from the LTSs configuration, deployment, and environment lead to transferability of T&E results to real-world scenarios. Despite their potential of increasing operational efficiency, transparency, and safety, the use of LTS in warehouse environments remains seldom. The *T&E 4Log Framework* aims to increase the transferability of T&E results to warehouse applications while fulfilling the stakeholders' T&E requirements for comparability, repeatability, feasibility, and comprehensibility. The issue of transferability of test results is addressed by integrating the application into the T&E, modeling the processes, the environment, and the localization requirements. The modular design of the *T&E 4Log Framework* integrates and extends the ISO/IEC 18305 and therefore ensures comparability, repeatability, and comprehensibility. The feasibility of the framework was successfully demonstrated at the facilities of the ITL by implementing the *ITL-Testbed*. T&E was accordingly performed on a reflector-based LiDAR system, a contour-based LiDAR system, and a UWB system. The requirements and potential use of the LTS for the applications *Automated Pallet Booking*, *Goods Tracking,* and *Autonomous Forklift Navigation* were subsequently discussed. The T&E process reveals the benefit in the use of methodical T&E approaches and the potential of an objective determination of the LTS's localization quality. Future work consists in the validation of the *Process & Environment Modeling*, the determination of environmental influences, and the extension of the *Application Evaluation*.

## 7. Acknowledgments

This work is supported by the TUHH $I^3$ Project funding.